 \documentclass[manuscript]{acmart}
 \usepackage{todonotes}
 
 \usepackage{amsfonts,amssymb}

\AtBeginDocument{%
  \providecommand\BibTeX{{%
    \normalfont B\kern-0.5em{\scshape i\kern-0.25em b}\kern-0.8em\TeX}}}

\begin{document}

\title{Virbo: Multimodal Multilingual Avatar Video Generation in Digital Marketing}


% \author{Juan Zhang}
% % \authornote{Both authors contributed equally to this research.}
% \email{zhangjuan14@300624.cn}
% \affiliation{%

\author{Juan Zhang}
\affiliation{%
  \institution{Wondershare}
  \city{Changsha}
  \country{China}}
\email{zhangjuan14@300624.cn}

\author{Jiahao Chen}
\affiliation{%
  \institution{Wondershare}
  \city{Changsha}
  \country{China}
}
\email{chenjh@300624.cn}

\author{Cheng Wang}
\affiliation{%
  \institution{Wondershare}
  \city{Changsha}
  \country{China}
}
\email{wangcheng@300624.cn}

\author{Zhiwang Yu}
\affiliation{%
  \institution{Hunan University}
  \city{Changsha}
  \country{China}
}
\email{yuzhiwang96@gmail.com}

\author{Tangquan Qi}
\affiliation{%
  \institution{Wondershare}
  \city{Changsha}
  \country{China}
}
\email{qitq@300624.cn}

\author{Can Liu}
\affiliation{%
  \institution{School of Creative Media, City University of Hong Kong}
  \city{Hong Kong}
  \country{China}
}
\email{canliu@cityu.edu.hk }

\author{Di Wu}
\affiliation{%
  \institution{Hunan University \& Wondershare}
  \city{Changsha}
  \country{China}
}
\email{dwu@hnu.edu.cn}


% video clipping, background segmentation and replacing, special effects addition, 
\begin{abstract}
 With the widespread popularity of internet celebrity marketing all over the world, short video production has gradually become a popular way of presenting products information. However, the traditional video production industry usually includes series of procedures as script writing, video filming in a professional studio, video clipping, special effects rendering, customized post-processing, and so forth. Not to mention that multilingual videos is not accessible for those who could not speak multilingual languages. These complicated procedures usually needs a professional team to complete, and this made short video production costly in both time and money. This paper presents an intelligent system that supports the automatic generation of talking avatar videos, namely Virbo. With simply a user-specified script, Virbo could use a deep generative model to generate a target talking videos. Meanwhile, the system also supports multimodal inputs to customize the video with specified face, specified voice and special effects. This system also integrated a multilingual customization module that supports generate multilingual talking avatar videos in a batch with hundreds of delicate templates and creative special effects. Through a series of user studies and demo tests, we found that Virbo can generate talking avatar videos that maintained a high quality of videos as those from a professional team while reducing the entire production costs significantly. This intelligent system will effectively promote the video production industry and facilitate the internet marketing neglecting of language barriers and cost challenges.
\end{abstract}


% \keywords{audio-driventalking avatar generation, the multimodal model for video generation, multilingual video generation, human avatar.}

\begin{teaserfigure}
  \includegraphics[width=\textwidth]{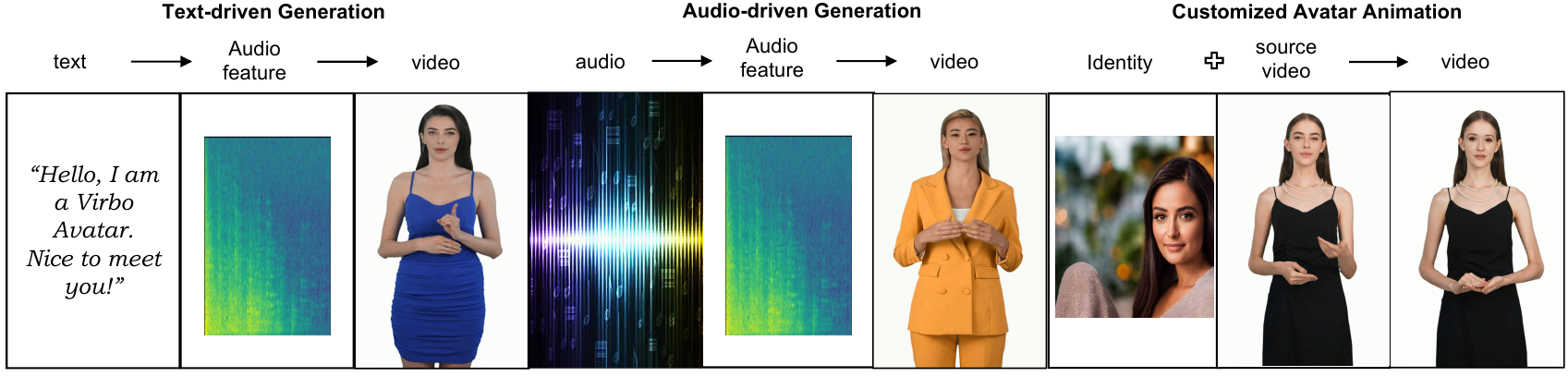}
  \caption{The Virbo system supports multimodal input such as text, audio, and images, to automatically generate customized talking avatar videos. This figure shows the customized generation of talking avatar demos based on user-specified multimodal data inputs.}
  \label{fig:teaser}
\end{teaserfigure}


\maketitle

\section{Introduction}
Short videos are a form of information presentation combining images and audio\cite{tan2021learning,guo2021vinci}. It can be used to promote products by describing product information and communicating intended benefits to the target audience. Creating a marketing short video can be difficult and laborious for video producers when facing overwhelming production processes such as creative script planning, character dubbing\cite{lahiri2021lipsync3d}, video special effects, and multilingual customization\cite{park2022synctalkface, zen2019libritts}. The most troublesome part of video production is model selection, model video filming, and post-production special effects. When filming videos of different scenes, such as sports, entertainment, leisure, etc., people need to choose different types of models to promote the corresponding products. If producers want to create short videos in different languages, they need to invite models from different languages. As we know, inviting models requires a significant amount of cost and time, not to mention shooting a large number of short videos in a short period of time\cite{ren2020Fast,zhang2023multilingual}. 

Despite that some techniques were proposed to automatically generate talking avatar videos to reduce the effort spent on model dubbing, they mostly focus on lip-sync accuracy with audio\cite{chen2021wavegrad,zhang2023metaportrait}. At the same time, they did not consider integrating multiple functions to help users create short videos more efficiently. The process of making short videos is beyond model dubbing. The video producer needs to select a proper model to shoot in the context of a given product scene. In order to achieve better visual effects, we also need to add post-production special effects to the video production. Besides, we need to customize multiple languages to promote videos in different countries and regions around the world\cite{liang2022expressive,song2022talking}. Our system can synthesize natural multilingual speech while maintaining the vocal identity of the speaker, as well as lip movements synchronized to the synthesized different language speech\cite{liu2021any}.

Recent advances in automatic video generation technology enable the task of talking avatar generation has drawn much attention due to its extensive real-world applications such as creating digital avatars\cite{chen2019hierarchical,wu2023styleme} and animated movies\cite{liu2022semantic}. However, to the best of our knowledge, no prior published study integrates the processes of talking face swapping\cite{chen2020simswap}, talking avatar generation, pose-production special effects, and multilingual customization into a whole framework, Three challenges need to be specifically addressed:
\begin{description}
\item[C1] How to prepare a large amount of short video digital human material, including facial photos of characters, talking avatar videos, and speaker audio. During the short video shooting process, we need a large number of models to shoot on-site, which will incur a lot of cost and time. If we can provide an intelligent generative framework that can generate marketing short videos for various scenarios by inputting specific images, videos, and audio of characters, it will greatly improve our efficiency in producing short videos.

\item[C2]
How to generate clear and realistic talking avatar videos? Current technologies generate talking avatar videos with unclear teeth, artifacts in the lip, and poor lip-sync accuracy. When we utilize artificial intelligence generative technologies to generate talking avatar videos, we should ensure that the character image needs to match the specific scene, the lip should be synchronized with audio and clear teeth. For example, when the audio is muted, the mouth shape of the same character should be unified as open or closed.

\item[C3] How to generate short videos in multiple languages. Short videos in multiple languages can promote faster dissemination of content in different countries and regions. In traditional short video production, we need to invite voice actors from different languages to dub, but it is laborious. An ideal solution is to generate short videos while specifying the language and ensuring the authenticity of the talking avatar.
\end{description}

To address these challenges, we introduce Virbo, an intelligent generation system for automatically generating multilingual multimodal-guided talking avatar short videos. Given a character image, talking video, and speaking audio, Virbo could generate a multilingual talking avatar short video with a specified personal identity video, personal identity voice, and input audio content. In order to provide users with more creative material choices when producing short videos, we provide a digital portrait library, voice library, and visual special effects for users to choose from. A deep generative model based on spatial deformation and feature decoder is employed to generate a vivid multimodal-guided talking avatar that has synchronized mouth shapes with input audio. Besides, Our Virbo supports multilingual talking avatar generation. Virbo incorporates a user interface from which users can easily produce talking avatar short videos to meet the user's production demand. The direct contributions and the novel aspects of this work include:
\begin{itemize}
    \item \textbf{Algorithm.} We propose an intelligent generative framework that can generate vivid multilingual multimodal-guided talking avatar short videos. The digital avatar we generate has synchronized mouth shapes with input audio, and the lip shape changes naturally and smoothly during the speech process. At the same time, our digital avatar supports the face swapping function. Last but not least, our generated digital avatar can speak in multiple languages like a native speaker.
    
    \item \textbf{System.} We introduce an intelligent short video creation system, 
    namely Virbo, to support the production process of multilingual talking avatar short videos. The system displays the specific digital portrait, voice, language, and generated talking avatar. At the same time, the system supports downloading and sharing of generated videos.
    
    \item \textbf{Evaluation.} 
    We demonstrate results from three forms of user studies: (1) a user study verifying the effectiveness of different components in the digital avatar generation model. (2) an efficiency and effectiveness test evaluating what factors could influence the generated digital avatar authenticity. (3) a feedback user study testing the functions of the Virbo system. The results show that Virbo can generate vivid multilingual multimodal-guided talking avatar short video as well as a native speaker. It also proves that our system can improve video producer's work efficiency. We also report feedback from professional video producers in an expert interview.
    
\end{itemize} 
% The rest of this paper is organized as follows. Section II summarizes the related works. Section III presents our group interview and system overview. Section IV describes the Virbo system in detail. Section V provides experimental results. Section VI gives further discussion and Section VII concludes our work.

\section{Related Work}
This section reviews some existing studies that are relevant to our research, including (1) audio-driven talking avatar generation, (2) multimodal model for video generation, and (3) multilingual video generation.

\subsection{Audio-driven talking avatar Video Generation}
The task of animating virtual humans from arbitrary speech sequences has drawn considerable attention, among which the talking avatar generation is especially significant.
Talking head generation aims to synthesize synchronized mouth shapes in every facial frame in a source video according to driving audio while keeping identity and head pose consistent with the source video frame, which includes  
 person-specific talking avatar generation and person-independent talking avatar generation\cite{stypulkowski2023diffused,chen2020talking}. 
 % One-shot talking face focus on driving one reference facial image with synchronic lip movements, and rhythmic head motions. The existing few-shottalking avatar generation makes great efforts to synthesize realistic faces.
 with the development of deep learning, a number of works generate person-specific high-quality results by leveraging the GAN-based pipelines. 
 Prajwal et al.\cite{prajwal2020lip} used a pre-trained discriminator that is already accurate at detecting lip-sync errors to inpaint the mouth areas. Ji et al.\cite{ji2021audio} proposed the cross-reconstructed emotion disentanglement technique to decompose speech into two decoupled spaces, a duration-independent emotion space and a duration-dependent content space. Lu et al.\cite{lu2021live} proposed Live Speech Portraits as the first audio-driven talking avatar animation system with photorealistic renderings in real-time. Thies et al.\cite{thies2020neural} presented a novel approach for audio-driven facial video synthesis by a deep neural network that employs a latent 3D face model space.
 
 Other researchers tend to seek a person-independent generating framework that can address all speaker identities. Zhou et al.\cite{zhou2019talking} enable arbitrary-subject talking face generation by learning disentangled audio-visual representation. Liang et al.\cite{liang2022expressive} proposed the granularly controlled audio-visual talking avatars, which controlled lip movements, head poses, and facial expressions of a talking avatar by decoupling the audio-visual driving sources through prior-based pre-processing designs. Song et al.\cite{song2022everybody} presented a method to edit target portrait footage by translating arbitrary source audio into arbitrary video output. 
 % However, it is not enough to drive thetalking avatar video generation just through audio.

\subsection{Multimodal-guided Talking Avatar Video Generation}
Multimodel-driven talking avatar generation refers to animating a portrait with the given pose, expression, and gaze transferred from the driving image and video\cite{zhang2023metaportrait,mcnutt2023design}. Recently, many attempts have achieved significant progress in these tasks\cite{ma2023styletalk,jang2023s}. Zhou et al.\cite{zhou2021pose} proposed a framework to generate pose-controllable talking faces by operating on non-aligned raw face images and an identity reference. Xu et al.\cite{xu2023multimodal} introduced a novel pipeline based on the multi-conditional diffusion model to afford complex texture and identity transfer, generating high-quality talking face generation for all driven modals. Yin et al.\cite{yin2022styleheat} proposed a unified framework based on a pre-trained StyleGAN that enabled high-resolution talking face generation, disentangled control by a driving video or audio, and flexible face editing. There are also methods utilizing 3D face model priors to provide a powerful tool for rendering and editing portrait images by parameters modulation. HeadGAN\cite{doukas2021headgan} pre-processed the 3d mash as input of the network to enable the model to operate as a real-time reenactment system. PIRender\cite{ren2021pirenderer} is proposed to enable photo-real editing of facial expressions, head rotations and translations with the parameters of 3DMMs based on disentangled modifications and efficient neural renderer. Zhong et al.\cite{zhong2023identity} proposed a two-stage framework consisting of a novel transformer-based landmark generator for audio-to-landmark generation and landmark-to-video rendering procedures.

There are also some multimodal large language models for visual understanding and generation\cite{hamalainen2023evaluating,dang2023choice}. Peng et al.\cite{peng2023kosmos} introduced a multimodal large language model, enabling new capabilities of perceiving object descriptions and grounding text to the visual world. Li et al.\cite{li2023videochat} initiated an exploration into video understanding by introducing VideoChat, an end-to-end chat-centric video understanding system. This work integrates video foundation models and large language models via a learnable neural interface, excelling in spatiotemporal reasoning, and video understanding\cite{li2023m}.

\subsection{Multilingual Talking Avatar Video Generation}
Expanding the task of synthesizing a talking avatar video to support multiple languages would significantly reduce the amount of effort required to widen the target audience to the global population. Recent works in talking face generation claim that their models support input speeches in any language\cite{lahiri2021lipsync3d,ma2023styletalk}. Lahiri et al.\cite{lahiri2021lipsync3d} presented a video-based learning framework for animating personalized 3D talking faces from audio. Maiti et al.\cite{maiti2020generating} presented a framework for bilingual text-to-speech which is able to transform a monolingual voice to speak a second language in a bilingual speaker embedding space while preserving speaker voice quality. FastSpeech\cite{ren2020fastspeech} can synthesize speech relying on an autoregressive teacher model for duration prediction and knowledge distillation in TTS. Shen et al.\cite{shen2018natural} described a neural network architecture composed of a recurrent sequence-to-sequence feature prediction network that mapped character embedding to mel-scale spectrograms for speech synthesis directly from text. Kong et al.\cite{kong2020hifi} modeled periodic patterns of audio for enhancing sample quality to achieve both efficient and high-fidelity speech synthesis. Chen et al.\cite{chen2021wavegrad} introduced WaveGrad2, a non-autoregressive generative model to estimate the gradient of the log conditional density of the waveform given a phoneme sequence. Kim et al.\cite{kim2021conditional} presented a parallel end-to-end TTS method that generated more nature-sounding audio by adopting variational inference augmented with normalizing flows and an adversarial training process. 

Besides speech synthesis-related studies, some works have focused on multilingual text-to-speech models with cross-lingual capabilities. Zhang et al.\cite{zhang2019learning} presented a multispeaker, multilingual text-to-speech synthesis model based on Tacotron that is able to produce high-quality speech in multiple languages. Nekvinda et al.\cite{nekvinda2020one} introduced an approach to multilingual speech synthesis that used the meta-learning concept of contextual parameter generation and produced natural-sounding multilingual speech using more languages.

\section{Group Interview and System Overview}
We aim to design Virbo that can automatically generate multilingual multimodal-guided talking avatar short videos given a text, a facial image, talking video, and speech audio. To better understand video producers' workflow and creative thinking, we conducted an in-depth group interview with four experienced video producers to investigate the general workflow of video production. Based on the interview feedback, we developed a user interface as shown in Fig.2 driven by our Virbo system. In this section, we describe the interview procedure, discuss the user feedback, and give a summary of the Virbo system.

\subsection{Group Interview}
We conducted a group interview with four video producers (three male and a female, aged 26-35) in an online way. All producers have three to five years of video editing and creation. We set up 10 questions on the relevant data collected and asked each producer to answer these questions according to their creative experience. We designed a question outline on the creative process from two topics: the production workflow of marketing short videos, and the difficulties they have when producing short videos. During the group interview, we asked the producers based on the question outline, responded to producers' queries in an interactive way, and recorded the producers' core insights. At the end of the group interview, we summarized the core insights based on the producers’ feedback on the two crucial topics as follows:

\subsubsection{The complicated production workflow of marketing short videos.} According to the interviewees, there are five common steps they followed when producing the marketing short video. First, design the style, write the script with specified duration, and target audience of a short video and outline the content and structure of the video. Second, select actors, determine marketing product props and scenes, actor lines, and visual effects. Third, shoot a speech video according to the script and requirements. Fourth, actor dubbing, editing materials, adding special effects, multilingual customization, and improving audio quality. Fifth, export the video to the appropriate format.

\subsubsection{The difficulties of producing short videos.}
All producers agreed that inviting models speaking specified multilingual languages and scheduling models' shooting times may be a challenge for video producers. One producer replied, "The cost of inviting models is too high, especially inviting foreign models". Another producer said that "It is impossible to shoot a short video in one shot, as models often forget their lines during the filming process". The producers also brought up another issue: a professional short video production team usually needs to hire staff who possess professional skills such as editing, multilingual dubbing, and special effects production. It is laborious, time-consuming and expensive to post-produce the short videos.

\subsection{System Overview}
Based on the feedback gathered from the interview, we have developed a user interface for automatically generating a set of talking avatar videos given the user-specified personal identity video, personal identity voice, language, inputting audio content, and video special effects. The Virbo system contains three major functionalities: generate user-specified faces and perform face swapping on the model, multimodal-guided talking avatar video generation, and multilingual talking avatar video generation. Here we present the system design goals extracted from our discussion with producers. Then, we illustrate how we achieve our design goals by introducing each system function.

\subsubsection{Design goals} 
Virbo is designed based on the following goals (G). Firstly, the Virbo is designed to reduce the labor cost of producers inviting models to shoot (G1). Secondly, Virbo is designed to alleviate the overload of models' work and reduce their shooting time and boring repeated dubbing work. Our system supports multimodal inputs to automatically generate batches of talking avatar videos (G2). Besides, Our system supports multilingual and multi-tone talking avatar video generation based on the user's selection, which could promote faster dissemination of short videos to different audiences in different countries and regions (G3). Last, our system provides a visual effects library for users to select, which could help them enrich the content of video creation (G4).

\subsubsection{Multimodal-guided talking avatar video generation} 
After loading the user interface as shown in Fig.2, the user can browse a series of models avatars in the system and select a model for video generation. During the data preprocessing, we invited the models to record a five-minute speech video in advance and used these videos as training data to generate talking avatar. Then user only needs to input a specified audio segment or text as the speech content for generating the talking avatar videos. If a user inputs a paragraph of text in any language, our system can translate it into the language chosen by users. Then our system can convert the text into a speech according to the text-to-speech Application Programming Interface (API). After the user completes the input, they can choose the speech tone, such as a mature British male accent or an American boy accent. If the user wants to specify the portrait of the model, they can upload a photo or facial image or use our AI-based character portrait generation interface to generate a facial image. 

% which include a feature encoder module, an alignment module, an adaptive affine transformation module, a feature decoder module, and a loss function module. 
% The encoder module extracts the features from the source image, driving audio, and reference images. The alignment module encodes the aligned information of the head pose between the source image and the reference image. The adaptive affine transformation module can deform feature maps with misaligned spatial layouts from fused features and reference image features. The decoder module is used to generate the dubbed image. The loss function module controls the synchronization of lips and driver audio.

These functions aim to help producers to reduce the work burden and labor costs. which is inspired by the second design difficulty as mentioned by our group interview. With this function, video producers can produce short videos more efficiently. Fig.4 gives an overview of the model architecture, which will be detailed in Section 4.1. It consists of two modules, which include a face swapping module and a talking avatar video generation module. The face swapping module is used to customize the portrait of character images. The talking avatar video generation module support generating a vivid talking avatar that synchronizes lips with audio. After preparing the relevant input files, click the generate button, and our system Virbo can generate a vivid talking avatar video. We will evaluate automatic talking avatar video generation with a user study in the scenario of producing short videos, so as to see whether this function can relieve video producers from the laborious work and generate realistic and high-definition talking avatar videos.

\subsubsection{Multilingual talking avatar video generation}
Considering the wide dissemination range of short videos, we need to increase our audience in different countries and regions. In the traditional video shooting process, the multilingual customization process occurs after video production. However, in our system, as long as we select the language to be translated before clicking the generate button, our system Virbo can effectively complete the multilingual customization process of talking avatar videos. This off-the-shelf function aims to help producers promote their video dissemination in different countries and regions, which is inspired by the third design difficulty as mentioned by our group interview. Fig.5 gives an overview of the multilingual voice cloning model architecture, consisting of two modules, which are a hybrid connectionist-temporal-classification-attention phoneme recognizer and a synthesis model with a mixture-of-logistic(MoL) location-relative attention module. A bottle-neck feature extractor (BNE) is obtained from the phoneme recognizer and is used to extract bottle-neck features (BNF) as spoken content representations of speech signals. The synthesis module incorporating speaker representations can generalize to unseen speakers. The benefit of doing so is that the seq2seqMoL approach can directly support any-to-many voice cloning.

\begin{figure}[tb]
\centering
  \includegraphics[width=0.8\columnwidth]{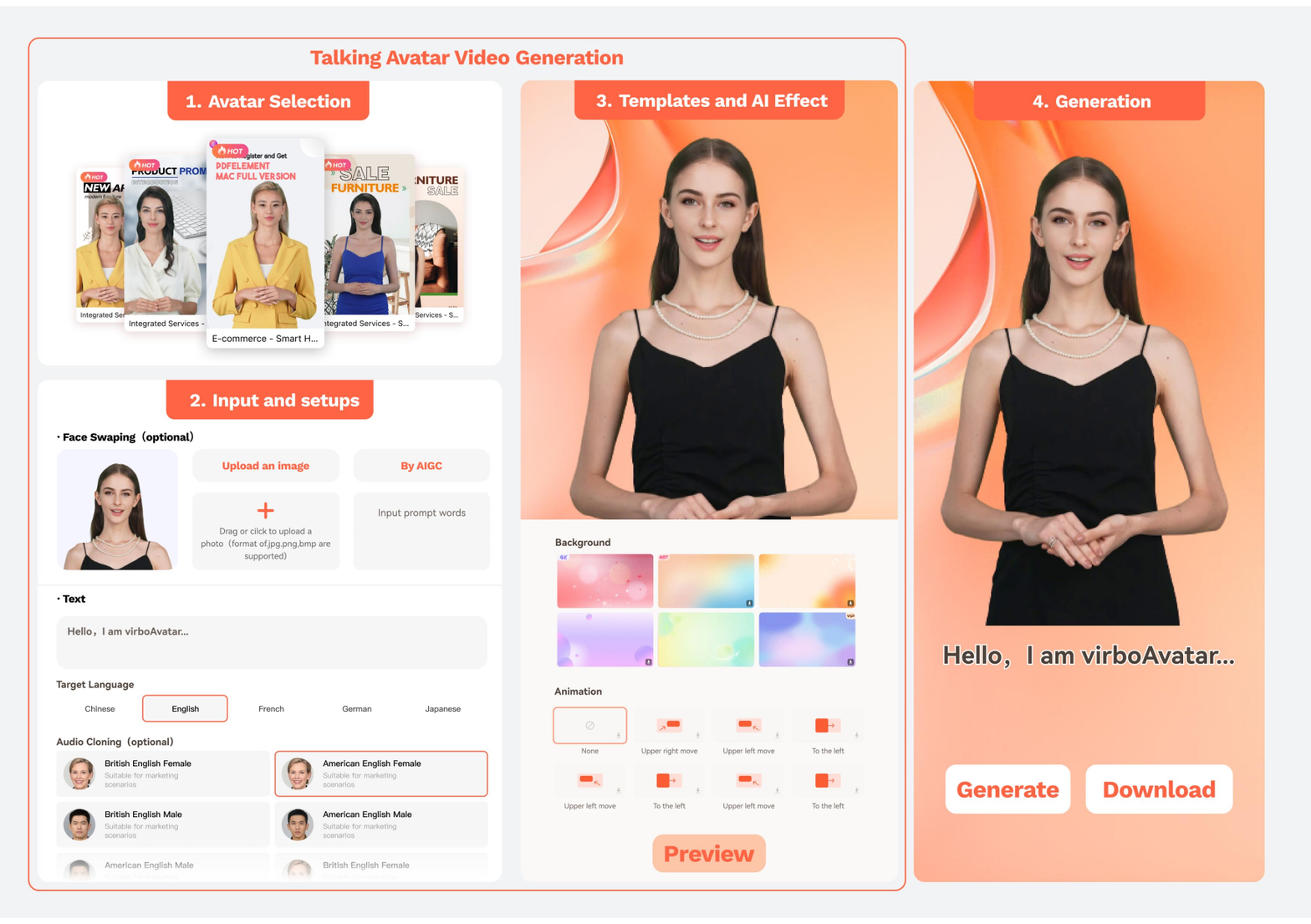}
  \caption{The user interface of the Virbo system consists of four parts, (1) Avatar selection area is used for users to browse the avatar template and determine video style; (2) The Input and setups area is used to support users to customize models' portraits, speech content and speaking language according to their uploading file; (3) Template and AI Effect Preview area is used to select background and animation visual effect as well as preview generated talking avatar videos; (4) Generation area is allowed for users to generate a complete talking avatar video and download.}
  \label{fig2}
\end{figure}

% \vspace{-10mm}
% Given the user-specified personal identity video, personal identity voice, inputting audio content, and video special effects, Virbo could use a deep generative model to generate a set of talking avatar videos
\section{System Description}
\begin{figure}[htbp]
\centering
  \includegraphics[width=1.0\columnwidth]{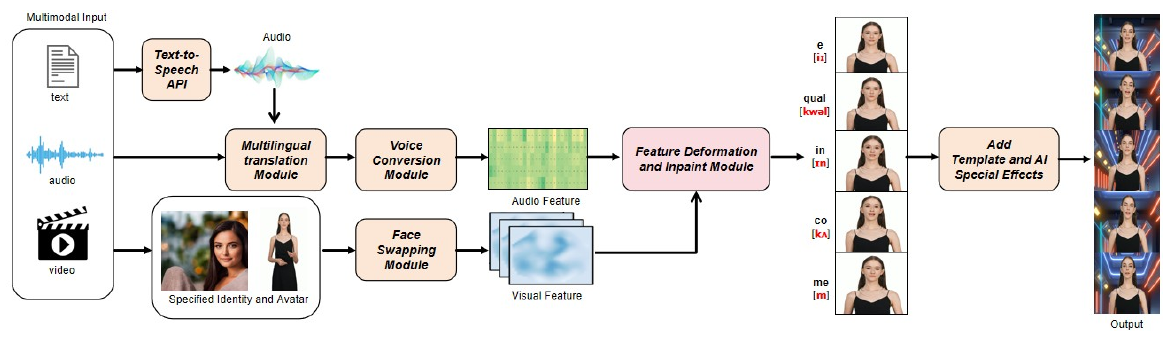}
  \caption{An overview of pipeline and architecture of Virbo system.}
  \label{fig3}
\end{figure}

As the Fig.~\ref{fig3} displayed, our system supports multilingual translation, voice cloning, face swapping, and talking avatar video generation. In the following subsection, we will introduce detailed technical details for each section.
\subsection{User Interface}
The user interface of Virbo is composed of two stages: the talking avatar video generation and the post-production video generation as shown in Fig.~\ref{fig2}. In the talking avatar generation stage, the user can browse a series of model templates and select a model as the basis for video generation. Then user needs to input a specified audio segment or text as the speech content for generating the talking avatar videos. Then our system can convert the text in a specified language into a speech according to the text-to-speech Application Programming Interface (API). After the user completes the input, they can choose sound speech and the tone of their speech, such as a mature British male accent or an American boy accent. If the user wants to specify the portrait of the model, they can upload a character portrait or use the character portrait generation interface to specify the generation of a character portrait. After preparing the relevant input files, click the preview button, and our system Virbo can generate a vivid talking avatar video. In the video post-production stage, if users need to add special effects to the generated video, they can select some subtitles and animation to the video, such as choosing subtitle styles such as font, color, size, and background animation for the background. Then users can download the completed videos and publish them on their social media platforms.

\subsection{voice cloning with Sequence-to-Sequence Model}
\subsubsection{Seq2seq phoneme recognizer}
This module focuses on developing an any-to-many voice cloning approach, which is expected to convert a source speech signal from an arbitrary speaker to a target speaker. Some published methods based on phonetic posteriorgrams (PPGs) are inaccurate in speech alignments which could result in inconsistent audio frames. To address these aforementioned deficiencies of approaches, we utilized a bottle-neck feature extractor with a seq2seq auto-regressive synthesis model. We first train an end-to-end hybrid connectionist-temporal-classification-attention (CTC-attention) phoneme recognizer. A bottle-neck feature extractor (BNE) is obtained from the phoneme recognizer and is used to extract bottle-neck features (BNF) as spoken content representations of speech signals. We then train a multi-speaker seq2seq BNF-to-spectral synthesis model which is equipped with a mixture-of-logistic (MoL) location-relative attention module, where each speaker is represented as a one-hot vector. 

In the phoneme recognizer, we adopt the CTC-attention model, CTC is a latent variable model that monotonically maps an input sequence to an output sequence of shorter length. By using conditional independence assumptions, the posterior distribution $p(Y|X)$ is factorized as follows:
\begin{equation}
\begin{aligned} 
P_{ctc}(Y|X)&=\sum_{z}\prod_{t}p(z_{t}|z_{t-1},Y)p(z_t|X)p(Y) \\
 P_{att}(Y|X)&=\prod_{l}p(y_{l}|y_{1},\cdots,y_{l-1};X),
\end{aligned} 
\label{eq1}
\end{equation}

where the frame-wise posterior distribution $p(z_t|X)$ is conditioned on all inputs $X$, we use a dynamic programming algorithm to compute the summation.

We regard the CTC objective as an auxiliary task to train the attention model encoder, which contains a VGG-Prenet and a BiLSTM encoder. we conduct utterance-level mean-variance normalization before feeding into the phoneme recognizer model. The training objective is to maximize a logarithmic linear combination of the CTC $p_{ctc}(Y|X)$ and attention $p_{att}(Y|X)$ 
 objectives.
\begin{equation}
\mathbb{J}_{seq2seq}=\lambda log P_{ctc}(Y|X)+(1-\lambda)log P_{att}(Y|X)
\label{seq2}
\end{equation}

\subsubsection{Seq2seqMoL synthesis and conversion procedure}
The synthesis model is an encoder-decoder model, where the encoder contains two networks, which include a bottle-neck feature encoder, and a pitch encoder, and the decoder contains attention-based RNN. The bottle-neck feature encoder contains two bidirectional GRU layers. The pitch encoder employs CNN structure, which operates continuously interpolated logarithmic $F0$ and unvoiced-voiced flags (UV) features input. The attention mechanism in the decoder is a location-relative extension.

Our approach computes the mel-spectrogram, continuous $Log-F0s$ and UV flags. Then the $Log-F0s$ are converted linearly in log-scale from the source to target speakers.
\begin{equation}
\mbox{Log-F0}_{vc}=\frac{\theta_{target}}{\theta_{source}}(\mbox{Log-F0}_{source}-\mu_{source})+\mu_{target}
\label{eq2}
\end{equation}
where $\mu$ and $\theta$ represent the mean and standard deviation of the log-scaled $F0$.
The bottle-neck features, $Log-F0_{vc}$ and $UV$ flags are added element-wise after going through the bottle-neck feature encoder and pitch encoder respectively. The output is concatenated with the target speaker embedding to form the encoder outputs. The MoL attention decoder then generates the converted mel-spectrogram from encoder outputs. A neural vocoder is finally used to generate waveform from the converted mel-spectrogram.

% We obtained a bottle-neck feature extractor from an hybrid CTC-attention phoneme recognizer.

\subsection{Face Swapping}
We utilized an efficient network, aimed for generalized and high fidelity face swapping. Previous approaches either lack the ability to generalize to arbitrary identity or fail to preserve attributes of the target face. like facial expression and gaze direction. To address these aforementioned deficiencies of approaches, we illustrated the ID injection module which transfers the identity information of the source face into the target face, so the relevance between the identity information and the weights of the decoder can be removed. Then a weak feature matching loss is used to constrain each attribute of the result image to match the target image so that the generated result can be aligned with the target image at the semantic level.

The ID injection module is composed of two parts, the identity extraction part and the embedding part. In the identity extraction part, we need a face recognition network to extract the identity vector from the source image. In the embedding part, we use the residual blocks to inject the identity information into the features. After the injection of identity information, we pass the modified features through the decoder to generate the final result.

In our architecture, the weak feature matching loss can be written as follows, $m$ is the layer where we start to calculate the weak feature matching loss.
\begin{equation}
L_{wFM}(D)=\sum_{i=m}^{M}\frac{1}{N_{i}} \Vert D^{(i)}(I_{R})-D^{(i)}(I_{T}) \Vert_{1}
\label{eqmatch}
\end{equation}

Our Loss function has four components, including Identity Loss, Reconstruction Loss, Adversarial Loss, and Weak Feature Matching Loss.

\begin{equation}
\begin{aligned} 
% P_{ctc}(Y|X)&=\sum_{z}\prod_{t}p(z_{t}|z_{t-1},Y)p(z_t|X)p(Y) \\
%  P_{att}(Y|X)&=\prod_{l}p(y_{l}|y_{1},\cdots,y_{l-1};X),\\
 L_{Id}&=1-\frac{v_{R}\cdot v_{S}}{\Vert v_{R} \Vert_{2}\Vert v_{S} \Vert_{2}}\\
L_{Rec}&=\Vert I_{R}-I_{T} \Vert_{1}\\
L_{wFMs}&=\sum_{i=1}^{2}L_{wFM}(D_i)
\end{aligned} 
\label{eqloss}
\end{equation}
The overall loss function can be written as:
\begin{equation}
\mathbb{L}=\lambda_{Id}L_{Id}+\lambda_{Rec}L_{Rec}+L_{adv}+\lambda_{wFM}L_{wFM_sum}
\label{eqlo}
\end{equation}

\subsection{Talking Avatar Video Generation}
To achieve photo-realistic talking avatar dubbing on high-resolution videos is a key challenge. To address this problem, we utilized a spatial deformation network and inpainting network for high-resolution talking avatar dubbing with rich textual details. The spatial deformation network consists of three feature encoders, an alignment encoder, and an adaptive affine transformation module. The spatial deformation network performs spatial deformation to create deformed feature maps encoding mouth shapes at each frame, to align with the input audio and the head poses of source images. The inpainting network consists of a feature decoder to generate adaptive mouth movements from the deformed feature and other attributes such as head pose and facial expressions from the source image.

 Given a source image, a driving audio, we utilize the deepspeech network as audio encoder to extract the audio feature $F_{audio}$ and use the Resnet as an image feature encoder to extract the source image feature $F_{s}$ and reference image feature $F_{r}$. Next, we concatenate the source feature and reference feature to input into one alignment encoder to compute alignment feature $F_{align}$. Finally, the audio feature and alignment feature are used to spatially deform the reference feature $F_{r}$ into the deform feature $F_{d}$. We utilize the adaptive affine transformation (AdaAT) operator to realize the spatial deformation. The  AdaAT can deform feature maps with misaligned spatial layouts from fused features and reference image features by feature channel-specific deformations. AdaAT operator computes different affine coefficients such as rotation, translation, and scaling in different feature channels by fully connected layers.
\begin{equation}
\left[
\begin{array}{c}
   \overline{x}_{c} \\
   \overline{y}_{c} 
\end{array}
\right]
=
\left[
\begin{array}{ccc}
    s^{c}cos(\theta^{c}) & s^{c}-sin(\theta^{c}) & t_x^{c} \\
    s^{c}sin(\theta^{c}) & s^{c}cos(\theta^{c}) & t_y^{c}
\end{array}
\right]
\left[
\begin{array}{c}
    x_c \\ 
    y_c \\
    1
\end{array}
\right]
\end{equation}

We utilized perception loss, GAN loss, and lip-sync loss to train our network. The perception loss is written as:
\begin{equation}
\mathbb{L}_{p}=\sum_{i=1}^{N}\frac{\Vert V_{i}(I_{o})-V_{i}(I_{r}) \Vert_{1}+\Vert V_{i}(\overline{I}_{o})-V_{i}(\overline{I}_{r})_{1}}{2NW_{i}H_{i}C_{i}}
\label{eqp}
\end{equation}
where $I_{o}$ represents dubbed image, $I_{r}$ represents the real image, $\overline{I}_{o}$ represents the downsample dubbed image, $\overline{I}_{r}$ represents the downsample real image. $V_{i}$ represents the $i_{th}$ layer in VGG-19 network.

The GAN loss is described as follows:
\begin{equation}
\begin{aligned} 
\mathbb{L}_{D}&=\frac{1}{2}E(D(I_{r})-1)^{2}+\frac{1}{2}E(D(I_{o})-0)^{2} \\
\mathbb{L}_{G}&=E(D(I_{o})-1)^{2}
\end{aligned} 
\label{eqGAN}
\end{equation}

We add a lip-sync loss to improve the synchronization of lip movements with input audio. The lip-sync loss is written as:
\begin{equation}
\begin{aligned} 
\mathbb{L}_{sync}=E(CLIP(A_{d},I_{o})-1)^{2}
\end{aligned} 
\label{eqlip}
\end{equation}

We sum the above losses as the final loss, which is written as:
\begin{equation}
\begin{aligned} 
\mathbb{L}=\lambda_{p}\mathbb{L}_{p}+\lambda_{sync} \mathbb{L}_{sync}+\mathbb{L}_{GAN}
\end{aligned} 
\label{eq4}
\end{equation}
% feature encoder module, an alignment module, an adaptive affine transformation module, a feature decoder module, and a loss function module. 
% The encoder module extracts the features from the source image, driving audio, and reference images. The alignment module encodes the aligned information of the head pose between the source image and the reference image. The adaptive affine transformation module can deform feature maps with misaligned spatial layouts from fused features and reference image features. The decoder module is used to generate the dubbed image. The loss function module controls the synchronization of lips and driver audio.

\section{Evaluation}
We evaluate the capability of our Virbo system in automatically generating talking avatar videos from three aspects. First, we conducted a user study on the impact of different functions on video production. 30 participants were invited to this study to test these functions to understand the participants' preferences. Second, we made an individual user study on the performance of Virbo by comparing the creation efficiency of video producers. Third, we made a user study to evaluate the performance of Virbo's functionality. In the procedure of user study, we recruited 10 participants to create some short videos. Then 100 participants were recruited to assess the quality of generated talking avatar videos. Last, we conducted a semi-structured interview with three professional short video producers to evaluate the user experience of Virbo. In the in-depth interview, the three experts were asked to use Virbo to create some short videos and provide some advice on the performance of Virbo. In this section, we first introduce the data preparation process for training the Virbo system. Then, we introduce the system function design, system evaluation, and discuss subjective feedback from participants regarding the corresponding experiments.

\subsection{Experiment 1: Performance of Model}
\begin{figure}[htbp]
\centering
\includegraphics[width=0.8\columnwidth]{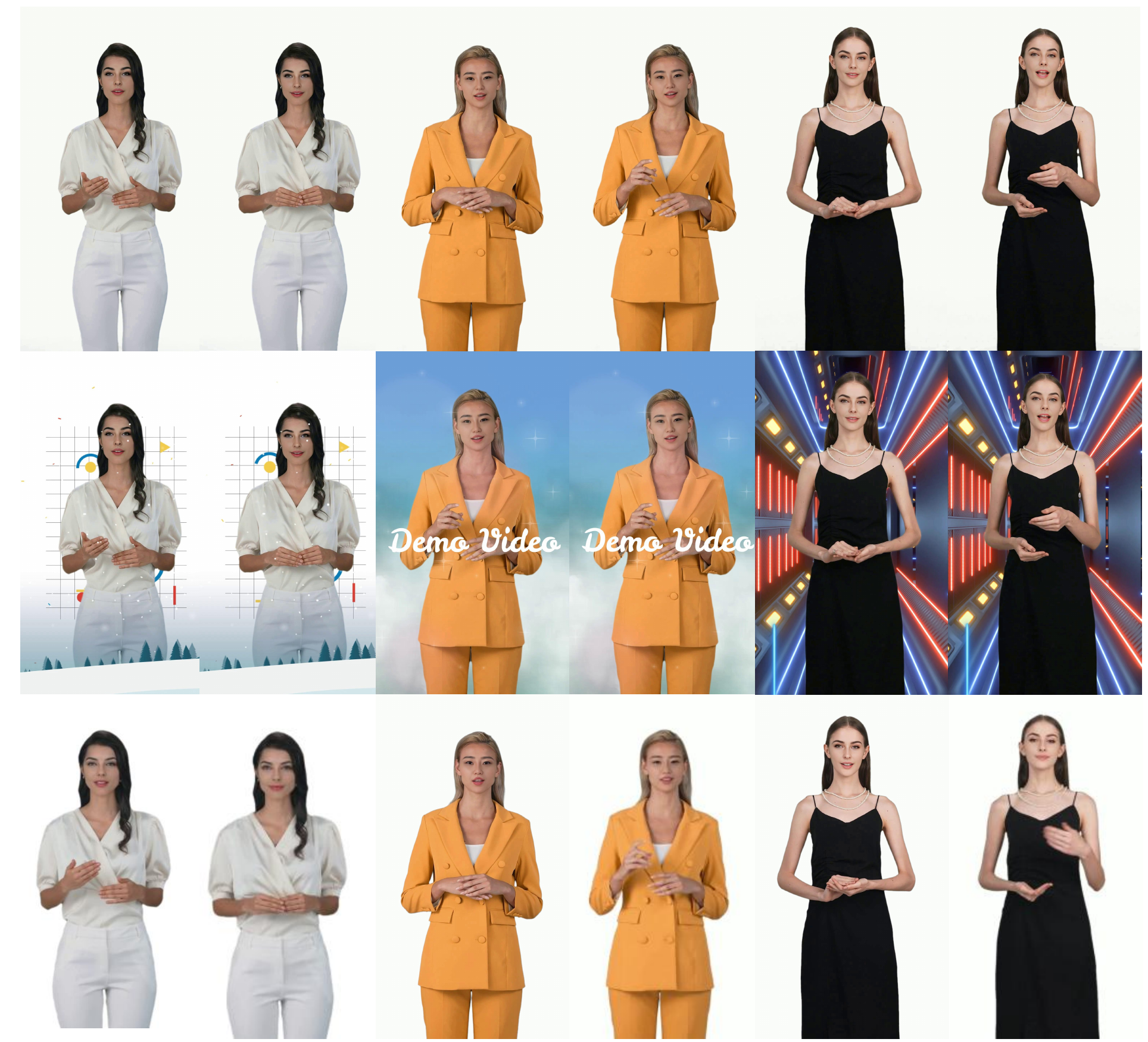}
  \caption{The frames of talking avatar video generations. The first-row graphics are from the Virbo system without visual effects, the second-row graphics are from the Virbo system with visual effects, and the third-row graphics are from another baseline system.}
  \label{fig4}
\end{figure}

We use four quantitative metrics to evaluate the experimental performance of our model. To evaluate the visual quality of generated digital talking avatars, we compute the metrics of Structural Similarity (SSIM), Peak Signal to Noise Ratio (PSNR), SSIM focuses more on the structural information of image content, while PSNR focuses on the peak signal-to-noise ratio and places greater emphasis on the accuracy of pixel values and the degree of image distortion. To evaluate the audio-visual synchronization, we compute the metrics of Lip Sync Error Distance (LSE-D) and Lip Sync Error Confidence (LSE-C). In the evaluation of visual quality, our system gets the best results on all metrics. A higher SSIM and PSNR represent better talking avatar image generation results. A lower LSR-D denotes a higher audio-visual match and the speech and lip movements are in synchronization. A higher LSE-C, the better audio-video correlation. Table~\ref{table:tab1} showed our system had better results in visual results and audio-video correlation compared to other baseline SenseTime. the Ground Truth (GT) videos were used to compare videos' audio-video correlation. Fig.~\ref{fig4} displayed the generated results from the Virbo system with or without visual effects, and another baseline. There is no function for the baseline to add visual special effects to video production. The talking avatars from the baseline method have lower accuracy in lip synchronization with audio and natural expression than our system.

% We assembled five hundred models to shoot the 5-minutes video to generate a dataset of 500 model template. 

\begin{table}[htbp]
\caption{The quantitative comparison to other methods.}
\label{table:tab1}
\centering
\begin{tabular}{ccccc}
\toprule  
	Methods&\multicolumn{4}{c}{\textbf{HDTF Dataset}~\cite{zhang2023dinet}}\\
 &\textbf{PSNR}~$\uparrow$ &\textbf{SSIM}~$\uparrow$ &\textbf{LSE-D}~$\downarrow$ &\textbf{LSE-C}~$\uparrow$\\
		\midrule  
            SenseTime~\cite{zhou2021pose}& 41.53 & 0.908 & 8.32 & 6.32 \\
	\textbf{Virbo}& 42.65 & 0.913 & 7.15 &7.53     \\
		GT & - & - & 6.89 & 5.19 \\
\bottomrule 
\end{tabular}
\end{table}

To run the follow-up user study, 20 participants (10 male and 10 female, mean age 25) are invited
to participate in this experiment by rating the talking avatar video in different user experience indicators. Each participant was invited to give his or her rating on video fluency, portrait clarity, mouth shape accuracy, and facial expression naturalness, which are crucial for evaluating the quality of the generated digital talking avatar. As described in Table.~\ref{table:tab2}, Our system virbo exceeds SenseTime method in all metrics.

\begin{table}[htbp]
\caption{User rating of talking avatar video generation.}
\label{table:tab2}
\centering
\begin{tabular}{ccccc}
\toprule  
	Methods &\textbf{Video fluency} &\textbf{Portrait clarity} &\textbf{Mouth shape accuracy} &\textbf{Expression naturalness}\\
		\midrule  
            SenseTime~\cite{zhou2021pose}& 7.7 & 8.0 & 6.5 & 6.8 \\
	\textbf{Virbo}& 7.9 & 8.3 & 6.8 &7.0     \\
\bottomrule 
\end{tabular}
\end{table}

\subsection{Experiment 2: Evaluation on Creation Time of Short video}
We also conducted a user study on the completion of shooting a short video by traditional video production and Virbo system-assisted video production to evaluate the performance of Virbo system by comparing the efficiency of traditional video production and Virbo system-assisted video production. The results are shown in Fig.~\ref{fig5}. In short video production, (a) actor speech shooting, (b) actor dubbing, (c) multilingual customization, and (d) special effects production are the most time-consuming aspects, which used different colors to represent the corresponding parts. The x-axis in the figure represents the generated video ID, $V1_V$ represents the first video generated using the virbo system. and $V1_t$ represents the first video generated using the traditional video production method. The y-axis in the figure represents the time cost on creating videos. We focus on these aspects to compare the production efficiency of traditional video production and Virbo system-assisted video production. We invited 20 video producers (10 females and 10 males, mean age of 27) to participate in video production. All participants were asked to create five short videos of marketing scenarios with a duration of less than thirty seconds. We provide six product props and models for participants to select. 

\subsubsection{Results.}
\begin{figure}[htbp]
\centering
\includegraphics[width=0.6\columnwidth]{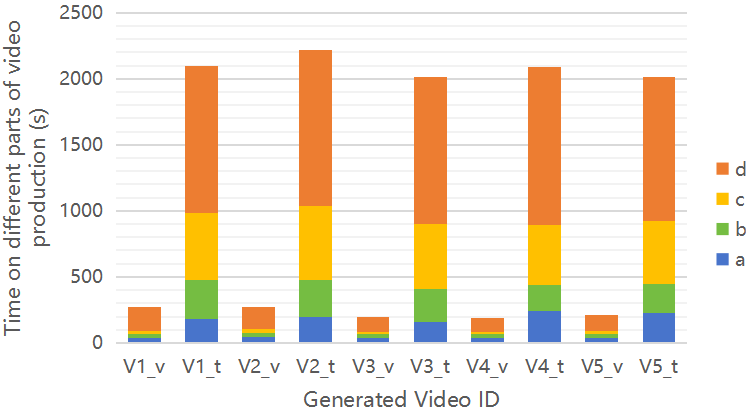}
  \caption{Task completion time for users using different creative methods}
  \label{fig5}
\end{figure}

We counted the completion time of (a) actor speech shooting, (b) actor dubbing, (c) multilingual customization, and (d) special effects production for each video producer to evaluate the performance of Virbo system-assisted video production. As shown in Fig.~\ref{fig5}, when participants create the first video, the average time of producers with the assistance of virbo system is about 224 seconds, while the average time of producers with the traditional production method is about 2046 seconds. As they continued to produce more short videos, the completion time of participants with the traditional production method depended on the models' proficiency in product introductions and the complexity of creating visual special effects. While the completion time of participants with the assistance of the Virbo system is almost consistent. 

In summary, the experiment results show that Virbo system can help video producers produce short videos in a shorter amount of time.

\subsection{Experiment 3: Efficiency and Effectiveness}
To evaluate the quality of generated talking avatar videos, we set up a group of experiments to evaluate whether our Virbo system has good efficiency and effectiveness in universality, personalization, and simplicity.

\subsubsection{Procedure.} 
We recruited two teams of participants, where one team consisted of 3 professional video producers, and the other team consisted of 3 amateur video producers. Each professional producer was asked to design 5 short videos based on different marketing scenarios by traditional video-producing methods, such as TikTok and YouTube. Each amateur producer was also asked to use the Virbo system to create corresponding 5 short videos. However, they were asked to imitate these videos created by professional video producers by the Virbo system. At last, we collected 5 videos
from each professional producer, and 5 videos from each amateur producer. Then, for each video, the participants were asked to identify whether it was produced by professional human producers or with the assistance of the Virbo system. Besides, all participants were asked to rate these three indicators including universality, personalization, and simplicity of the Virbo system on a 10-point Likert scale from "very poor" to "very good".

\subsubsection{Hypotheses.}
We have pilot research with 6 participants to use Virbo to generate talking avatar videos. According to the pilot research, we have two aspects of hypotheses as follows:
\begin{itemize}
\item{H1:} In the aspect of information dissemination and visual effects, the impact of short videos produced by the Virbo system on the audience should be similar to those generated by professional video producers. 
\item{H2:} In terms of system function, the virbo's function are universal, personalized and simple compared with the other production method. 
\end{itemize}

% simplicity, and personalization
\subsubsection{Results.}

We compared the support rate of the videos by two teams. The professional team had the best performance which was rated 8.7, while those videos generated by our system were rated 8.1. These results showed that our system has a good performance on talking avatar authenticity and lip synchronization with audio, which verified that our first hypothesis $H1$ that in the aspect of information dissemination and visual effects, the impact of short videos produced by the Virbo system on the audience should be similar to those generated by professional video producers. As the Table.~\ref{table:tab4} described, our system rating exceeds the baseline by a significant margin in three indicators. It also verified our second hypothesis H2 that Virbo has better system functions compared with the other methods.

\begin{table}[htbp]
\caption{User rating of Virbo System in different dimensions}
\label{table:tab4}
\centering
\begin{tabular}{cccc}
\toprule  
	Methods &\textbf{Universality} &\textbf{Personaliztion} &\textbf{Simplicity} \\
		\midrule  
            SenseTime~\cite{zhou2021pose}& 6.2 & 7.1 & 6.3 \\
	\textbf{Virbo}& 7.3 & 8.3 & 7.1     \\
\bottomrule 
\end{tabular}
\end{table}

\subsection{Expert Interview}
To further evaluate our system, we conducted an in-depth interview with three professional designers, we named P1, P2, and P3. Both of them had over 5 years of video production experience. Before the interview, we briefly introduced the Virbo system design and demonstrated the usability of the system. After trying out the Virbo system for several video generations, we carried out the interview on main three topics: the quality and realism of video generation, the personalized function of Virbo system, and the universality of the Virbo system. The experts were also asked to provide some comments on the performance of Virbo. We described some of the comments below.

\subsubsection{The efficiency of creating short videos.} The three professional video production practitioners expressed surprise and appreciation for the efficiency of Virbo's video production.   P1 mentioned:" The Virbo system provides vivid and photo-realistic talking avatar video generation, and the video generation speed is very fast. It only takes about one minute to generate a 30-second video. Even for complex post-production, it only costs two minutes." P2 said:" It can generate different speech language versions, just by selecting the corresponding language. This has saved me a lot of time in video production and allowed me to spend more time on creative video scripts. Due to my overseas business, I often need to produce videos in different language version in traditional methods, which is time-consuming and laborious." P3 was impressed with the authenticity of talking avatar generated by Virbo:" The talking avatar is photo-realistic, and it has a good lip synchronization with speaking audio, continuity in mouth shape changes, and natural facial expressions of characters. The generated digital talking avatar has effectively replaced the work of real human in marketing or knowledge training scenarios."

\subsubsection{System.}
All interviewers believed that Virbo was an effective tool and its user interface meets the principle of simplicity and efficiency
P1 said:" Virbo system provided a large number of model templates, which greatly relieved our workload of selecting models. Moreover, the Virbo has classified these templates into different scenes, which has inspired us to determine the shooting theme." P2 was impressed with the number of digital model templates and the types of visual special effects:" The Virbo system has a variety of personalized functions, providing users with diverse templates and materials to create rich short videos. With the help of the virbo system, I can make short videos more efficiently." P3 thought that the function of multilingual voice convertion was useful, "As long as choose the language and speech voice, I can easily generate multilingual digital talking avatar videos, which greatly accelerates the promotion of videos in different countries and regions and greatly helps my marketing business." P1 also said:" The visual effects are great, providing a richer display format for videos, which is essential for those who pursue novelty." Overall, all interviewers felt that Virbo was a powerful and effective system, but there is still potential for improvement. P1 suggested the generated digital talking avatars are all in front view, and if the generated video can switch between multiple shots, such as alternating far and near, it could be more in line with the habit of live action. P2 raised his concerns that :"could the generation of a large number of digital talking avatars lead to the abuse of portrait copyright?" and he said that:" In the rapid development process of generative AI, we also need to specify corresponding management standards to avoid the disasters it brings."

\section{Discussion}
As suggested in our user experiment, Virbo is designed to generate a vivid and photo-realistic talking avatar video close to the level of professional video producers. Through user customization, our Virbo system can generate different voices, languages, portraits, and visual special effects. In this section, we discuss the limitations and implications of our work.

\subsection{Edge Cases and Failure Modes}
\begin{figure}[tb]
\centering
  \includegraphics[width=0.7\columnwidth]{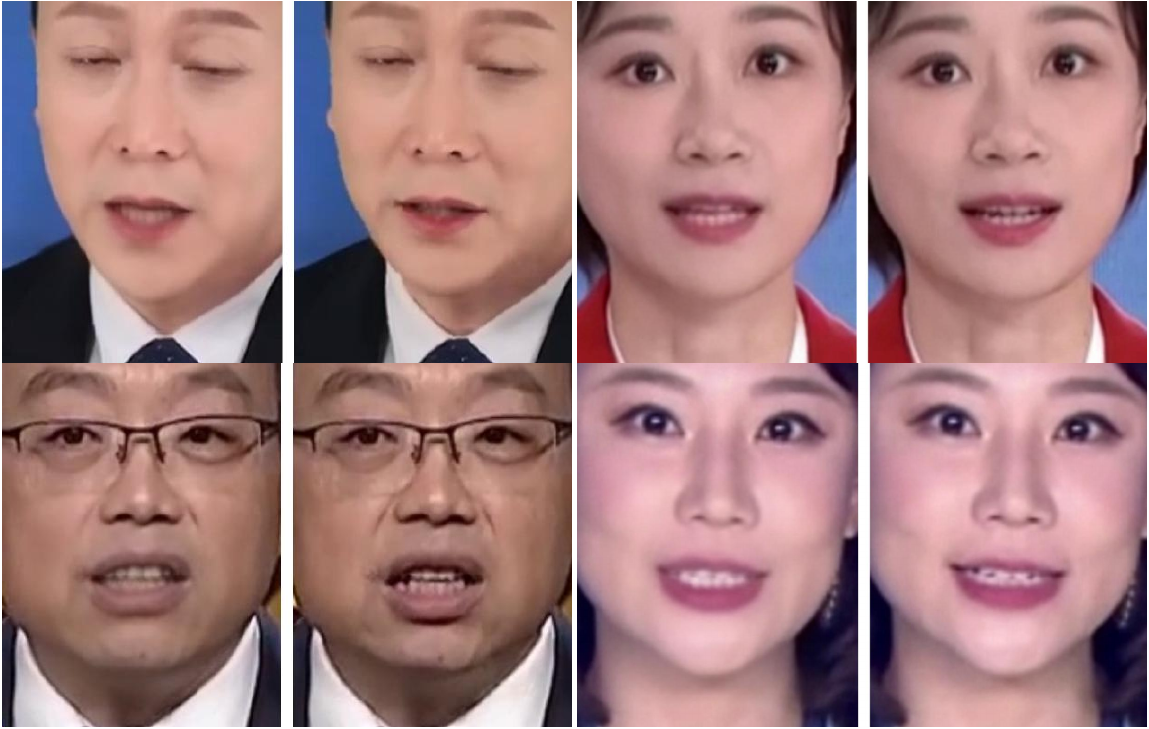}
  \caption{Edge cases in Virbo system.}
  \label{fig_bad}
\end{figure}

In the current implementation of Virbo system, though it can generate the talking avatar videos with comparably better results than existing works in most cases as illustrated in previous experiments, there are still some defective edge cases. Fig.~\ref{fig_bad}. gives several cases where the system does not work well, where the generated image represents the frame of the generated videos, and the Ground Truth image represents the original videos. When the user uploads facial images with poor image quality, the generated digital person's facial clarity is poor. When producing pronunciation phonemes with small changes in mouth shape, the generated digital talking avatar mouth shape is inaccurate. When speaking too fast, the generated teeth may have artifacts or become unclear. In the future, we can add some weak supervision to generate better mouth shape changes and tooth clarity for digital talking avatars.

\subsection{Other Limitations}
In the following part, we discuss several limitations of our system that are identified from the experiments and user feedback.

\subsubsection{Data Preparation.} 
Model materials also need to collect more character speech videos to cover more scenes. At the same time, we need to classify model materials for different scenes and provide detailed usage instructions, because which types of models are more suitable for promoting which categories of products and used in which scenes are valuable marketing experiences. At the same time, we need to collect some rich emotional speech, so that the tone of the model changes with emotional fluctuations, in order to provide the audience with a more realistic experience. In the future, we need to enrich the attributes of our collected data, allowing our digital talking avatars to have more flexible and customized functions, and providing users with a more vivid and realistic experience.

\subsubsection{User Interface.} In the file inputting area, we can only support image, text, and audio file input, but we can not support video files. In future work, we will support direct editing of videos uploaded by users, allowing them to use their portraits as models. Meanwhile, we need to provide more customized function interfaces to facilitate the creation of diverse and rich content videos for users. These functions can help users produce videos more efficiently.

\subsubsection{Function Diversity.} Our system can only support face swapping and voice cloning, but some users need to change the model's clothing and hairstyle. How to change the local details of the model to create richer video content is a future work to need improvement. Besides, we need talking avatars to speak with rich emotions. In order to make digital talking avatars more realistic, the users hope the tone of the talking avatars' voices can change with emotions. It could not be better if the emotions of the talking avatars' voices were consistent with facial expressions.

% \subsection{Traditional short video production tools vs. AIGC assisted short video production tools}

% \subsection{Generalizability}
% As we have seen in both the quantitative and qualitative analysis, Virbo is favored by a majority of users in their
% produce process. In fact, our system can be generalized to other digital avatar application scenarios such as Metaverse.
% For example, if there is a large number of shoe sketches and image datasets,

\section{CONCLUSION AND FUTURE WORK}
In this work, we present an intelligent talking avatar video generation system, Virbo. To the best of our knowledge, it is the first integrated and off-the-shelf intelligent system with diverse personalized functions and a large number of digital talking avatar templates and visual special effects templates. It consists of many practical functions such as multilingual customization, voice cloning, face swapping, talking avatar dubbing, and visual special effects rendering. Given the user-specified personal identity video, personal identity voice, inputting audio content, and visual special effects, Virbo could generate a set of talking avatar videos with customized voice, language, and portrait. The objective experimental metrics showed the talking avatar video generated by Virbo are more photo-realistic and has better lip synchronization with talking audio compared with other baseline system. The subjective user studies indicated that our system Virbo is rated higher regarding accuracy of mouth shape pronunciation, naturalness of facial expressions, and authenticity of portrait image. Besides, our system can achieve video creation that is comparable to the level of professional creation. Future work is to overcome the current limitation on enriching the speaker's voice emotions and facial expressions.
.

% 

% \begin{table}
%   \caption{Frequency of Special Characters}
%   \label{tab:freq}
%   \begin{tabular}{ccl}
%     \toprule
%     Non-English or Math&Frequency&Comments\\
%     \midrule
%     \O & 1 in 1,000& For Swedish names\\
%     $\pi$ & 1 in 5& Common in math\\
%     \$ & 4 in 5 & Used in business\\
%     $\Psi^2_1$ & 1 in 40,000& Unexplained usage\\
%   \bottomrule
% \end{tabular}
% \end{table}

% \begin{table*}
%   \caption{Some Typical Commands}
%   \label{tab:commands}
%   \begin{tabular}{ccl}
%     \toprule
%     Command &A Number & Comments\\
%     \midrule
%     \texttt{{\char'134}author} & 100& Author \\
%     \texttt{{\char'134}table}& 300 & For tables\\
%     \texttt{{\char'134}table*}& 400& For wider tables\\
%     \bottomrule
%   \end{tabular}
% \end{table*}

% \begin{figure}[h]
%   \centering
%   \includegraphics[width=\linewidth]{sample-franklin}
%   \caption{1907 Franklin Model D roadster. Photograph by Harris \&
%     Ewing, Inc. [Public domain], via Wikimedia
%     Commons. (\url{https://goo.gl/VLCRBB}).}
%   \Description{A woman and a girl in white dresses sit in an open car.}
% \end{figure}

\bibliographystyle{ACM-Reference-Format}
\bibliography{main}

\end{document}